\documentclass[aps,prl,twocolumn]{revtex4}

\usepackage{graphicx}

\newcommand{\Ec}{E_c }

\begin{document}

\title{Coherent Dynamics of a Josephson Charge Qubit}

\author{T.~Duty}
\email[]{tim@fy.chalmers.se}
\author{D.~Gunnarsson}
\author{K.~Bladh}
\author{P.~Delsing}
\affiliation{Microtechnology and Nanoscience, MC2, Chalmers
University of Technology, S-412 96 G{\"o}teborg, Sweden}

\date{\today}

\begin{abstract}
We have fabricated a Josephson charge qubit by capacitively
coupling a single-Cooper-pair box (SCB) to an electrometer based
upon a single-electron transistor configured for radio-frequency
readout (RF-SET). Charge quantization of $2e$ is observed and
microwave spectroscopy is used to extract the Josephson and
charging energies of the box. We perform coherent manipulation of
the SCB by using very fast DC pulses and observe quantum
oscillations in time of the charge that persist to $\simeq10ns$.
The observed contrast of the oscillations is high and agrees with
that expected from the finite $E_J/E_C$ ratio and finite rise-time
of the DC pulses. In addition, we are able to demonstrate nearly
$100\%$ initial charge state polarization. We also present a
method to determine the relaxation time $T_1$ when it is shorter
than the measurement time $T_{meas}$.
\end{abstract}


\maketitle

\graphicspath{{..}}

Although a large number of physical systems have been suggested as
potential implementations of qubits, solid state systems are
attractive in that they offer a realistic possibility of scaling
to a large number of interacting qubits. Recently there has been
considerable experimental progress using superconducting
microelectronic circuits to construct artificial two-level
systems. A variety of relative Josephson and Coloumb energy scales
have been used to construct qubits based upon a single-Cooper-pair
box \cite{Nakamura99,Vion02} and flux qubits based upon a
3-junction loop \cite{vanderWal00,Chiorescu03}. Coherence times of
the order of 0.5 microseconds have been achieved for a
single-Cooper-pair box qubit\cite{Vion02}. Rabi oscillations
between energy levels of a single large tunnel junction have also
been observed \cite{Yu02, Martinis02}. Despite the encouraging
results, one aspect that is not well understood concerns the
contrast of the oscillations, which in all previously reported
experiments is smaller than expected.

The experimental systems reported so far can also be distinguished
by the readout method and the manner of performing single qubit
rotations. The first demonstration of coherent control of a
single-Cooper-pair box (SCB) \cite{Nakamura99} employed a weakly
coupled probe junction to determine the charge on the island. In
the more recent experiment reported by Vion \emph{et
al.}\cite{Vion02}, the SCB was incorporated into a loop containing
a large tunnel junction, for which the switching current depends
on the state of the SCB. Switching current measurements of SQUIDS
have also been used for flux and phase-type
qubits\cite{vanderWal00, Chiorescu03, Yu02, Martinis02}. Nakamura
\emph{et al.}\cite{Nakamura99} performed single qubit rotations by
applying very fast DC pulses to a gate lead in order to quickly
move the SCB into, and away from the charge degeneracy point. This
technique produces qubit rotations with an operation time that can
be of the order of the natural oscillation period. Other
experiments utilize microwave pulses to perform NMR-like rotations
of the qubit \cite{Vion02, Yu02, Martinis02,Chiorescu03}. The
latter approach requires less stringent microwave engineering,
since RF-rotations can be accomplished with pulses that are more
than order of magnitude slower in rise time and duration than the
natural oscillation period. By using fast pulses, however, single
qubit rotations can be performed approximately 20 times faster
than with RF-rotations.

%

%
\begin{figure}[tb]
    \centering
    \includegraphics[width = 0.8\columnwidth]{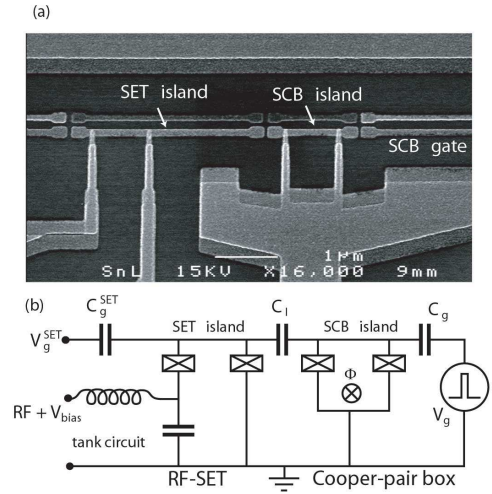}
    \caption{(a) Scanning electron micrograph of a sample. The device consists
             of a Cooper-pair box and SET electrometer and was fabricated
             from an aluminum (lighter regions) evaporated onto an oxidized Si substrate
             (darker regions). (b) Circuit diagram of box and electrometer.}
    \label{sample}
\end{figure}

In this letter, we report measurements made on a SCB-type qubit
with very fast DC pulses used to effect the qubit rotations, as in
Nakamura \emph{et al.}\cite{Nakamura99}. For our qubit, however,
the readout system consists of a single-electron-transistor (SET)
capacitively coupled to the SCB and configured for radio-frequency
readout (RF-SET). By incorporating a SET into a tank circuit,
RF-SET electrometers can be made that are both
fast\cite{SchoelkopfSci98} and sensitive\cite{AassimeAPL01}, and
are well-suited for SCB-qubit
readout\cite{AassimePRL01,Lehnert02}. One advantage of using
RF-SET readout is that it can easily be turned on and off,
although for the measurements reported here it is operated in a
continuous\cite{Lehnert02}. More significanly, the RF-SET is
fundamentally different from qubit readout based on swithing
currents in that it involves a weak measurement of the charge in
contrast to a yes or no answer. This could be a significant
advantage for readout of multiple qubit systems. For example, a
switching current measurement on one qubit would also directly
affect the other qubits due to its strong interaction.
Furthermore, switching current measurements are inherently
stochastic in time---an important consideration if one is
interested in measuring time correlations.

Electron beam lithography and double-angle shadow evaporation of
aluminum films onto an oxidized silicon substrate were used to
fabricate the combined SCB-SET system (see Fig. 1)\cite{Bladh02}.
The SCB box consists of a low-capacitance superconducting island
connected to a superconducting reservoir by two parallel junctions
that define a low-inductance SQUID loop. An additional gate lead
placed near the island is used to change the electrostatic
potential of the island by a gate voltage $V_g$ through a gate
capacitance $C_g$. By adjusting the magnetic flux $\Phi$ through
the loop, the effective Josephson energy is tunable as
$E_J=E_J^{max}|cos(\pi\Phi/\Phi_0)|$, where $\Phi_0$ is the
magnetic flux quantum ($h/2e$). Coupling of the SCB to the SET was
accomplished by extending part of the SET island to the proximity
of the box island and resulted in a weak dimensionless coupling,
$C_I/C_\Sigma$ of $0.5-4\%$ for samples with slightly different
geometries and total SCB capacitances $C_\Sigma$. $C_I$ is the
capacitance between the SCB and SET islands.

The samples were placed at the mixing chamber of a dilution
refrigerator with a base temperature of approximately 20mK. An
external superconducting magnet was used to produce a large field
parallel to the plane of the samples and a small superconducting
coil close to the sample was used to produce a field perpendicular
to the SQUID loop. This allowed us to independently suppress the
superconducting gap of the aluminum film, and change the effective
Josephson energy of the Cooper-pair box. All control lines were
filtered by a combination of low-pass and stainless steel and
copper-powder filters.  To present sharp pulses to the SCB gate, a
high frequency coaxial line was used having $10$dB attenuation at
$4.2$K and $20$dB at $1.5$K. The tank circuit had a resonant
frequency of $380$MHz and the RF-SET electrometer was biased near
a feature in the $IV$ versus $V_g^{SET}$ landscape known as the
double Josephson quasi-particle peak (DJQP) \cite{ClerkPRL02},
such that the bias current through the SET was typically
200-300pA. Under these conditions, the electrometer sensitivity
for the combined SCB-SET system was found to be 50
$\times10^{-6}e/\sqrt{\textrm{Hz}}$ with a bandwidth of 10MHz.
\begin{figure}[tb]
    \centering
    \includegraphics[width = 1.0\columnwidth]{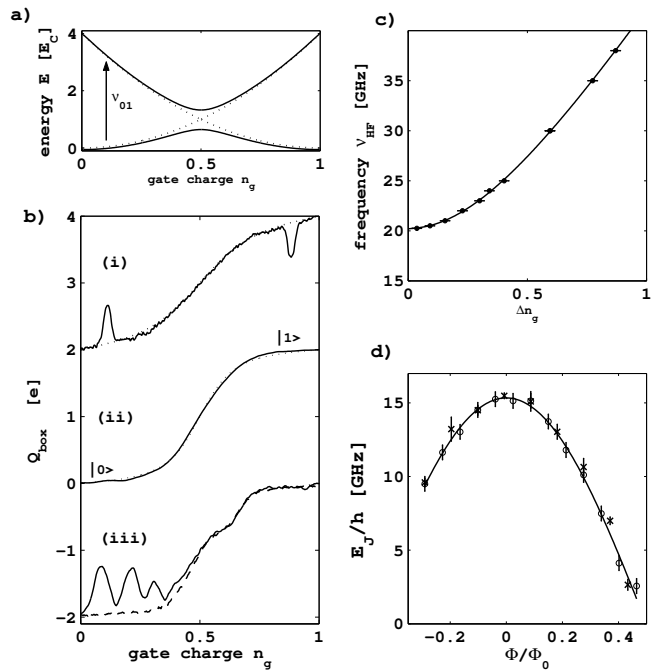}
    \caption{\textbf{(a)} Ground state and first excited state energies versus $n_g$ for
             $E_J/E_C=.7$ (solid line) and $E_J=0$ (dotted line).
             \textbf{(b)} Coulomb staircases, $Q_{box}$ versus $n_g$ for a sample
             with $\Ec/h=9.25\pm0.10$GHz and $E_J^{max}/h=20.24\pm0.10$GHz. In
             \textbf{(i)} a staircase taken from a sample under microwave
             radiation with a frequency of $35$GHz is shown. In
             \textbf{(ii)}, we show data where $E_J/h$ has
             been suppressed to $11.4$GHz by applying a small magnetic
             field through the SQUID loop.  The dotted lines in  \textbf{(i)}
             and  \textbf{(ii)} show the calculated
             $Q_{box}$ vs $n_g$ for the respective Josephson
             energies.  \textbf{(iii)} Measured $Q_{box}$ vs $n_g$ (solid line)
             when a fast pulse train is applied to the SCB gate
             having an amplitude $0.88e$, width $\Delta t=130$ps and repetition
             time $T_R=130$ns (here $E_J/h=4$GHz). The dashed line is the measured $Q_{box}$
             without the pulse train.  \textbf{(c)} Avoided level crossing shown by the
             positions of the spectral peaks and dips, $\Delta
             n_g=(n_g^{dip}-n_g^{peak})/2$,
             for varying microwave frequency.
             \textbf{(d)} Josephson energy $E_J$
             versus applied flux
             $\Phi/\Phi_0$ through the SQUID loop determined from
             spectroscopic data (circles) and coherent oscillation
             data (crosses). This sample had $\Ec/h=17.45\pm0.13$GHz
             and $E_J^{max}/h=15.5\pm0.2$GHz.}
     \label{fig:elevels_staircases}
\end{figure}

To make an artificial two-level system, the energy scales of the
SCB must be chosen so that $k_BT < E_C,E_J > \Delta$, where $E_C$
is the energy cost required to add a single electron to the island
and is set by total capacitance of the island $E_C=e^2/2C_\Sigma$.
If $\Delta$ is large enough compared to $E_C$, then the ground
states will be even parity states that differ only by the average
number of Cooper-pairs on the island. The effective Hamiltonian of
the box, including the Josephson coupling, is given by
\begin{equation}
\label{hamiltonian}
  H = 4E_C\sum_{n} (n-n_g)^2 |n\rangle\langle n | - \frac{E_J}{2}\sum_{n}(|n+1\rangle\langle n | + |n\rangle\langle
n+1
    |),
\end{equation}
where we define gate charge as $n_g=C_g V_g/2e-n_0$, and $n_0$ is
the offset charge due to stray charges near the box. Fig. 2(a)
shows the ground state and first excited state energy bands
calculated using the Hamiltonian in Eqn.(\ref{hamiltonian}). On
account of the $2e$ periodicity of the system, we limit the
discussion to gate charge $0\leq n_g \leq 1$.

In thermodynamic equilibrium, the actual quantity that determines
the island parity is the even-odd free energy difference
$\tilde{\Delta}(T)$, which differs from $\Delta$ due to entropic
considerations.\cite{Lafarge93}. Measurements were made for
several different samples with Coulomb enegies $E_C/k_B(E_C/h)$
ranging from 0.43 to 1.65$K$ (9-34GHz) and $E_J/E_C$ ratios of 0.4
to 2.25. Coloumb staircases were measured by slowly ramping the
box gate charge (at a rate of $\sim 40e/s$), and measuring the
power reflected from the RF-SET tank circuit. Examples of the
measured box charge $Q_{box}(n_g)$ is shown in Fig. 2b. Our
samples show $2e$ periodicity when $\Ec/k_B<1K$, well below the
gap $\Delta/k_B=2.2$K. For samples with larger $\Ec$, the
staircases acquire an extra step around $n_g=0.5$ due to poisoning
by non-equilibrium quasiparticles.

The characteristic energies $E_{C}$ and $E_{J}^{max}$ of the SCB
can be determined directly using microwave
spectroscopy\cite{Lehnert02}. When monochromatic microwaves are
applied to the SCB gate, resonant peaks and dips occur in the
measured $Q_{box}$ vs $n_g$ staircase when the microwave photon
energy matches the energy level splitting $\nu_{01}$ (see Fig.
2(b)). As the microwave frequency is reduced, the positions of
these peaks exhibit an avoided level crossing. Within the
two-level approximation, the positions of the single-photon
resonances depend on the applied frequency $\nu$ as $\Delta
n_g=\sqrt{h^2\nu^2-E_J^2}/8E_C,$ where we used $\Delta
n_g\equiv(n_g^{dip}-n_g^{peak})/2$. An example of such an avoided
level crossing is shown in Fig 2(c) where we have used
least-squares fits to this equation to estimate the Coloumb and
Josephson energies.


Measurements of time-resolved, coherent oscillations of the charge
on the SCB are made by slowly ramping the SCB gate charge $n_g$ as
before, while applying fast (non-adiabatic) rectangular DC-voltage
pulses of amplitude $a$ in the form of a pulse train to the SCB
gate. When $n_g$ is far away from the charge degeneracy, and $a$
such that $n_g+a$ places the system at the charge degeneracy, the
leading and trailing edges of each pulse act as successive
Hadamard transformations. In between, the system coherently
oscillates between charge eigenstates for a time $\Delta t$. The
trailing edge of the pulse returns the gate charge to $n_g$ with
the system in a superposition of ground and excited states
corresponding to the phase of the oscillation acquired during
$\Delta t$. After the pulse, the excited state component decays
with a relaxation time $T_1$. The pulses are separated by a
repetition time $T_R$, typically greater than $T_1$.

Since the staircase is acquired by ramping $n_g$ on a timescale
much slower than $T_R$, one measures an enhanced charge $\Delta
Q_{box}=Q_{box}^{on}-Q_{box}^{off}$, that is time-averaged over
the pulse train and proportional to the probability of finding the
system in the excited state. Peaks in $\Delta Q_{box}$ vs $n_g$
will be located at gate charges $n_g^{peak}$ such that
approximately a half-integer number of oscillations occur at gate
charge $n_g^{peak}+a$ during the time $\Delta t$. Fig. 2(b)
(bottom curve) shows a measured staircase using such a pulse
train. When one measures such staircases for varying $\Delta t$,
then time-domain oscillations are evident in the $\Delta t$
cross-sections of $\Delta Q_{box}$ vs $n_g$. This is shown in Fig.
4(a), where we plot the time evolution near the charge degeneracy
point. As $T_R$ is increased, the pulse-induced charge contributes
less to the time average. A simple calculation which ignores
coherence effects remaining after a time $T_R$ leads to the
dependence
\begin{equation}
\label{t1dep} \Delta Q_{box}(T_R) =
2n_0\frac{T_1}{T_R}\frac{1-e^{-T_R/T_1}}{1+e^{-T_R/T_1}},
\end{equation}
where $n_0$ is the initial peak amplitude. One can estimate $T_1$
at the readout gate voltage by measuring the dependence of the
peak amplitude on $T_R$ and fitting the data to this formula.

\begin{figure}[tb]
    \centering
    \includegraphics[width = 1.0\columnwidth]{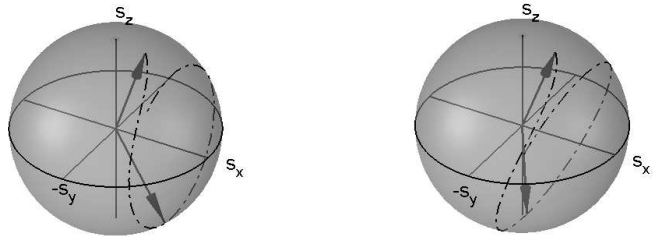}
    \caption{
    \textbf{(a)} Bloch sphere showing evolution about the charge degeneracy
    for $E_J/h=6$GHz, $E_C/h=9$GHz and a pulse rise-time of $35$ps.
    \textbf{(b)} Evolution for a pulse that takes the system past
    the charge degeneracy and achieves $100\%$ charge state polarization (downward pointing arrow).
    \label{fig:bloch}}
\end{figure}

Using such a pulse-train allows a measurement of the excited state
probability even when the relaxation time $T_1$ is shorter than
the measurement time. Population averaging and mixing limit the
maximum observed amplitude of the $\Delta Q_{box}$ oscillations to
$1e$ rather than $2e$, and this occurs only for an ideal
rectangular pulse, short repetition times and vanishing $E_J/E_C$
ratio. We have numerically integrated the Schr\"{o}dinger equation
using pulses with a finite risetime to study the various factors
affecting both the contrast of the oscillations and the degree of
charge state polarization. The results are best understood when
plotted on the Bloch sphere as shown in Fig. 3. The dash-dotted
lines show the evolution in time of the state vector starting from
an initial condition (upward pointing arrows) and evolving through
a maximum polarization of the charge state (downward arrows). The
initial state differs from pure $\textbf{S}_z$ due to a finite
$E_J/E_C$. Fig. 3(a) shows evolution for a pulse that takes the
system to the charge degeneracy. One finds that a finite pulse
rise-time mimics a higher $E_J/E_C$ ratio and reduces the
oscillation contrast. Nonetheless, Fig. 3(b) shows that by pulsing
past the charge degeneracy, one can achieve nearly $100\%$
polarization.

Measurements of the $T_R$ dependence of $\Delta Q_{box}$ are shown
in Fig. 4(b), where we have used an appropriate pulse amplitude
and duration to achieve nearly $100\%$ initial polarization. We
find $T_1=108\pm20$ns at the readout point $n_g\simeq0.25$ using a
least-squares fit to Eqn.(\ref{t1dep}). The data of Fig. 4(a)
shows an oscillation contrast of approximately $0.55e$. Taking in
to account the experimentally determined $E_J/E_C$ and $T_1/T_R$,
along with the measured risetime $\delta t \simeq 35$ps of our
pulse generator (Anritsu MP1763C), we find good agreement between
the observed and expected initial oscillation contrast. Numerical
solutions of Schr\"{o}dinger equation show that the reduction of
contrast due to the pulse risetime depends strongly on $E_J$. This
was checked experimentally by varying $E_J$ and found to be in
agreement with the calculations. A fit of the data of Fig. 4(a) to
an exponentially decaying sinusoid gives a oscillation frequency
$\nu_0=3.6$GHz and decoherence time of 2.9ns.

Using fast DC pulse trains, we were able to observe oscillations
in time out to 10ns. The dependence of decoherence time on gate
charge at the top of the pulse is shown in Fig. 4(c) and has a
sharp maximum at the charge degeneracy. This tells us that low
frequency charge noise is the dominant source of dephasing in our
qubit. As discussed and demonstrated by Vion \emph{et
al.}\cite{Vion02}, dephasing due to charge fluctuations is minimum
at the charge degeneracy. The coherence time found by Vion
\emph{et al.} is much larger than that found in our measurements
although their initial oscillation amplitude is much smaller. If
we extrapolate the gate charge dependence of $T_1$ (not shown) to
the charge degeneracy, we expect $T_1\sim10$ns. This implies that
the decoherence time we measure at the charge degeneracy is
dominated by the short relaxation time rather than pure dephasing.

Our measured $T_1$ was also found to be independent of SET bias
current in the sub-gap regime. Previously, a considerably larger
$T_1$ was measured in a similar sample using a spectroscopic
technique\cite{Lehnert02}, and found to agree with the theoretical
$T_1\simeq 1 \mu s$ expected from quantum fluctuations of a
$50\Omega$ environment. The reduced $T_1$ observed here could be
attributed to several factors. One is coupling to the relatively
unfiltered high frequency coaxial line, in addition to unwanted
coupling to other metal traces and the microwave environment of
our sample. In practice it is difficult to estimate the actual
real part of the impedance on these leads at such high transition
frequencies. Another possibility concerns quasiparticles. Our
samples were fabricated without quasiparticle traps and despite
the observed static $2e$-periodicity, there could be considerable
non-equilibrium quasiparticle dynamics occuring on these time
scales. Finally, it may be that the use of fast DC pulses leaves
the bath of charge fluctuators in a configuration having
additional channels for relaxtion of the qubit. Clearly what is
needed is $T_1$ measurements using a variety of
techniues---RF-rotations, spectroscopy, as well as fast
DC-pulse---on the same sample.

\begin{figure}[tb]
    \centering
    \includegraphics[width = 1.0\columnwidth]{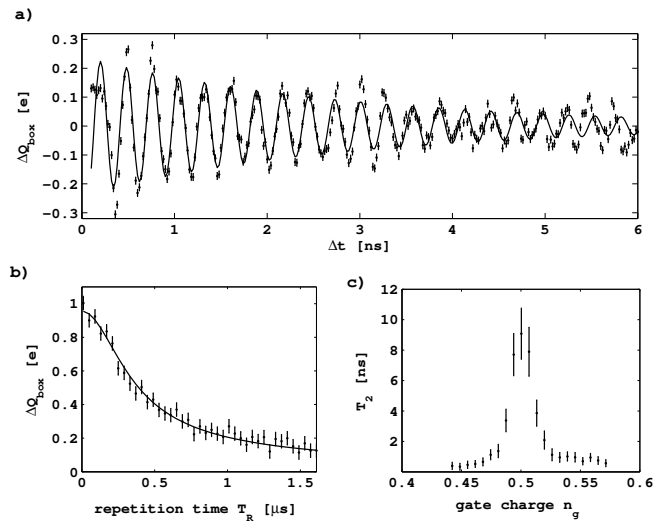}
    \caption{
    \textbf{(a)} $\Delta Q_{box}$ versus $\Delta t$. The solid line is a least-squares fit
     to an exponentially decaying sinusoid and gives a decay time $T_2\simeq 2.9$ns
     and frequency $\nu_0=3.6$GHz.  \textbf{(b)} Dependence
    of the peak height $\Delta Q_{box}$ on
    repetition time $T_R$ for $\sim100\%$ initial polarization.
    The solid line is a least-squares fit to Eqn.(\ref{t1dep}) giving
    $T_1=108\pm20$ns.
    \textbf{(c)} Decoherence time $T_2$ versus gate charge $n_g$.
    \label{fig:coosc1}}
\end{figure}

In conclusion, we have fabricated and measured a solid state qubit
based upon a SCB combined with a RF-SET readout system. Due to a
relatively short $T_1$, continuous measurement of the SCB was
employed. Fast DC pulses were used to coherently manipulate the
qubit and time-coherent oscillations of the charge were observed.
The initial contrast of the oscillations is relatively large and
quantitatively understood as due to finite $E_J/E_C$ combined with
the finite pulse risetime. Furthermore, nearly $100\%$ charge
state polarization can be achieved. The oscillations show a
maximum decoherence time at the charge degeneracy which indicates
that charge fluctuations dominate the dephasing rate.

\begin{acknowledgments}
We would like to acknowledge fruitful discussions with R.
Schoelkopf, K. Lehnert, A. Wallraff, G. Johansson, A K{\"a}ck, G.
Wendin and Y. Nakamura. The samples were made at the MC2
cleanroom. The work was supported by the Swedish SSF and VR, by
the Wallenberg and G{\"o}ran Gustafsson foundations and by the EU
under the IST-SQUBIT programme.

\end{acknowledgments}

\end{document}